\documentclass{ws-procs9x6}

\newcommand{\refeq}[1]{(\ref{#1})}
\def\etal {{\it et al.}}

\begin{document}

\title{TEST OF LORENTZ SYMMETRY BY USING\\
       A $^{3}$HE/$^{129}$XE CO-MAGNETOMETER}

\author{K.\ TULLNEY,$^*$ C.\ GEMMEL, W.\ HEIL, S.\ KARPUK, K.\ LENZ, YU.\ SOBOLEV }

\address{Institut f\"ur Physik, Universit\"at Mainz, Staudingerweg 7\\
55099 Mainz, Germany\\
$^*$E-mail: tullnek@uni-mainz.de}

\author{M.\ BURGHOFF, W.\ KILIAN, S.\ KNAPPE-GR\"UNEBERG, W.\ M\"ULLER,\\ A.\ SCHNABEL, F.\ SEIFERT, L.\ TRAHMS}

\address{Physikalisch-Technische-Bundesanstalt, Abbestrasse 2--12\\
10587 Berlin, Germany}

\author{U.\ SCHMIDT}

\address{Physikalisches Institut, Universit\"at Heidelberg, Philosophenweg 12\\
69120 Heidelberg, Germany}

\begin{abstract}
To test Lorentz symmetry we used a $^3$He/$^{129}$Xe co-magnetometer. We will give a short summary of our experimental setup and the results of our latest measurements. We obtained preliminary results for the equatorial component of the background field interacting with the spin of the bound neutron: $\tilde{b}^{n}_{\bot}<3.72\times10^{-32}$ GeV (95\,\% CL).
\end{abstract}

\bodymatter
\section{Introduction} 

According to the Standard-Model Extension (SME) described in Ref.\ \refcite{kostelecky}, Lorentz violating sidereal variations of the Larmor frequency of co-located spin polarized gases occur due to the rotation of the laboratory reference frame with respect to a relic background field. To search for this effect we measured the free spin precession Larmor frequency $\omega_{\mathrm{L}}=\gamma \cdot B$ of co-located $^3$He and $^{129}$Xe in a homogenous magnetic field \textit{B}, where $\gamma$ is the gyromagnetic ratio of the corresponding gas species.
Referring to the Cramer-Rao lower bound\cite{CRLB}, the upper limit for the frequency sensitivity of such a damped sinusoidal signal with white gaussian noise is given by
\begin{equation}
\sigma_{\mathrm{\nu}} \: \geq \: \frac{\sqrt{12}}{(2\pi)\cdot SNR \cdot \sqrt{\nu_{\rm BW}} \cdot T^{3/2}} \cdot \sqrt{C(T,T_{2}^{*})} ~,
\label{aba:crlb}
\end{equation}
where \textit{SNR} is the signal-to-noise ratio, $\nu_{\mathrm{BW}} = r_{\mathrm s}/2$ is the sampling rate $r_{\mathrm{s}}$ limited bandwidth, and \textit{T} the observation time. $C(T,T_{2}^{*})$ describes the effect of exponential damping with the transverse spin relaxation time $T_{2}^{*}$, whereby $C(T,T_{2}^{*}) \approx 1$ for observation times $T \leq T_{2}^{*}$. In the regime of motional narrowing and at low magnetic fields ($B \approx 1 \, \mathrm{\mu}$T) the $T_{2}^{*}$ for polarized spin samples can be in the order of days\cite{cates}. For a \textit{SNR} of 10000:1 in a bandwidth of 1 Hz and $T_{2}^{*} \approx 1$\,day, a sensitivity in frequency estimation of $\sigma_{\mathrm{\nu}} \approx 3$\,pHz can be reached\cite{ourPaper}. 

\section{Experimental setup}

\begin{figure}
\begin{center}
\psfig{file=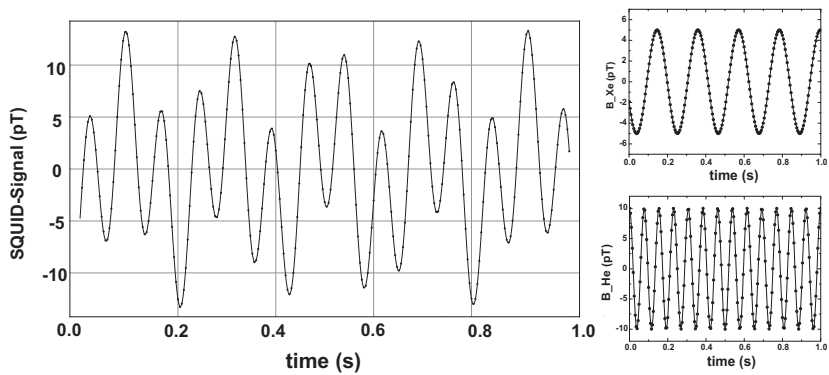, width=4.3in}
\caption{Left: time interval of recorded SQUID signal. Right: precession signals for $^3$He (above) and $^{129}$Xe (below), separated by a digital filter (bandwidth 1\,Hz) around 13.4\,Hz and 4.9\,Hz, respectively.}
\label{aba:signals}
\end{center}
\end{figure}

\vskip -1 pt
Our experiments were performed in the magnetically shielded room BMSR-2 at the PTB in Berlin. Its 7 layer $\mu$-metal shield preserves a low residual magnetic field of $<\,$2\,nT and a field gradient around the sample cell  $<\,$0.3\,pT/cm. A homogenous magnetic guiding field of $\approx$\,400\,nT was provided by two square coil pairs arranged perpendicular to each other in order to manipulate the sample spins. The maximal field gradients were $\approx$\,33\,pT/cm. Our sample cells were filled with a gas mixture of $^{3}$He, $^{129}$Xe, and N$_{2}$, whereby N$_{2}$ is needed as buffer gas for $^{129}$Xe\cite{chann}. The cells were placed directly below the Dewar that contains 300 LT$_{\mathrm{C}}$-SQUIDs which detect a sinusoidal \textit{B} field change caused by the spin precession of the gas atoms. 
The recorded signal is a superposition of the $^3$He and $^{129}$Xe precession 
signals which can be separated by applying, e.g., a digital filter (see Fig.\ \ref{aba:signals}). 
Each signal is sinusoidally modulated and decays exponentially with its respective relaxation time $T_{2, \mathrm{He(Xe)}}^{*}$. 

\vskip -15 pt

\section{Analysis}\label{aba:sec1}

For practical reasons, we did not analyze the Larmor frequency $\omega_{\mathrm {L,He(Xe)}}$ of $^3$He and $^{129}$Xe, but its equivalent, the phase $\Phi_{\mathrm{L,He(Xe)}}$. In order to cancel the dependence on ambient magnetic field fluctuations, we analyzed the weighted phase difference of the co-located precessing spin ensembles,
\begin{equation}
\Delta\Phi(t)=\Phi_{\mathrm{L,He}}(t)-\frac{\gamma_{\mathrm{He}}}{\gamma_{\mathrm{Xe}}}\cdot\Phi_{\mathrm{L,Xe}}(t)~.
\label{aba:domega}
\end{equation}
If there were no effects on the spin precession frequency, respectively the phase, the weighted phase difference $\Delta\Phi(t)$ is expected to be constant. 

\begin{figure}
\begin{center}
\psfig{file=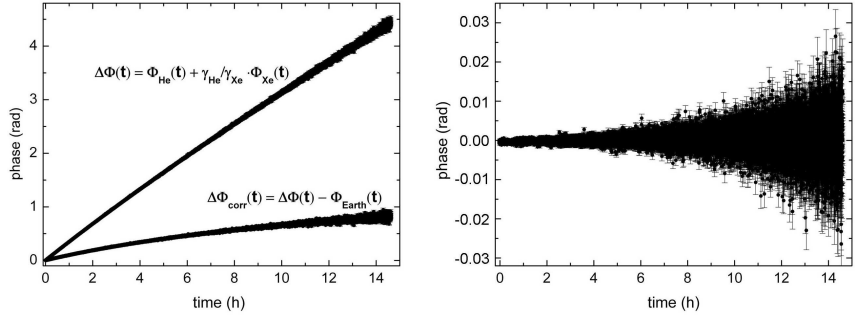}
\caption{Left: weighted phase difference before ($\Delta\Phi$) and after ($\Delta\Phi_{\mathrm{corr}}$) the subtraction of the Earth's rotation term ($\Phi_{\mathrm{earth}}$). Right: phase residuals for a decisive subrun.}
\label{aba:phases_residuals}
\end{center}
\end{figure} 
  
Figure \ref{aba:phases_residuals} shows the time dependence of $\Delta \Phi(t)$ which is mainly dominated by a linear term. Similar to the Foucault pendulum, the rotation of the SQUIDs because of the Earth's rotation with respect to the precessing spins causes a constant frequency shift $\omega_{\mathrm{earth}}=0.69 \cdot 10^{-4}$\,rad/s, and thus a linear term in the measured phase $\Phi_{\mathrm{earth}}(t)=\omega_{\mathrm{earth}}\cdot t$. If we subtract this linear term from the weighted phase difference, we get the corrected phase $\Delta \Phi_{\mathrm{corr}}(t)$ (see Fig.\ \ref{aba:phases_residuals}). In March 2009, we performed a measurement consisting of 7 long-term subruns in series, each with a duration of 10\,hours at least. For all subruns j, with $\mathrm {j}={1,...,7}$, we determined the weighted phase difference $\Delta\Phi(t)^{\mathrm{(j)}}$ and subtracted the phase shift $\Phi_{\mathrm{earth}}(t)$ in order to obtain $\Delta\Phi_{\mathrm{corr}}(t)^{\mathrm{(j)}}$. 
The latter one can be described by
\begin{eqnarray}
   \Delta \Phi_{\mathrm{corr}}^{\mathrm{(j)}}(t) &=& \Phi_\mathrm{0}^{\mathrm{(j)}} + \Delta\omega_{\mathrm{lin}}^{\mathrm{(j)}} (t-t_{\mathrm{0,j}}) \vspace{0.8cm} \notag \\
                       & & + E_{\rm He}^{\mathrm{(j)}} \exp\left(\frac{-(t-t_{\mathrm{0,j}})}{T_{\mathrm{2,He}}^{*\mathrm{(j)}}}\right)
                           + E_{\rm Xe}^{\mathrm{(j)}} \exp\left(\frac{-(t-t_{\mathrm{0,j}})}{T_{\mathrm{2,Xe}}^{*\mathrm{(j)}}}\right)~.
\label{aba:modellorentzfit}
\end{eqnarray}
Here, $t_{\mathrm{0,j}}$ is the starting time of each subrun with respect to the first subrun, $\Phi_\mathrm{0}^{\mathrm{(j)}}$ a phase offset, and $\Delta\omega_{\mathrm{lin}}^{\mathrm{(j)}} (t-t_{\mathrm{0,j}})$ a linear phase shift mainly arising from deviations of $\gamma_{\mathrm{He(Xe)}}$ from their literature values due to chemical shifts. $E_{\mathrm{He(Xe)}}^{\mathrm{(j)}}$ are the amplitudes of two exponential terms that reflect the respective phase shift due to coupled spin ensembles based on demagnetization fields in a non-ideally spherical cell. This phase shift is directly correlated to the decay of the signal amplitude of the precessing $^{3}$He and $^{128}$Xe spins\cite{ourPaper}. Since the $T_{2}^{*\mathrm{(j)}}$ times of helium and xenon can be determined directly from the experiment, four fit parameters for each subrun are left, i.e., the fit model is basically a linear function in the parameter set. By fitting the corrected phase difference $\Delta \Phi_{\mathrm{corr}}^{\mathrm{(j)}}(t)$ and subtracting the fit from the data we obtain the phase residuals, shown in Fig.\ \ref{aba:phases_residuals} for a decisive subrun. Because of the exponential decay of the $^3$He and $^{129}$Xe signal amplitudes --- mainly $^{129}$Xe with $T_{2}^{*}$ of only 4 to 5\,h --- the SNR decreases, respectively the phase noise increases, in time.

For a combined fit to all 7 subruns we defined a piecewise fit function
\begin{equation}
   \Delta \Phi_{\mathrm{corr}}\mathrm{(t)} = \sum^{7}_{\mathrm{j}=1} \Delta\Phi_{\mathrm{corr}}^{\mathrm (\mathrm j)}(t)  +  \Phi_{\mathrm{LV}} \cdot \sin(\Omega_{\mathrm{E}} t + \phi)~,
\label{aba:f=fi+sin+cos}
\end{equation}
where $\Phi_{\mathrm{LV}}$ is the amplitude of a possible Lorentz violating phase shift modulated with the frequency $\Omega_{\mathrm{E}}\approx2\pi/(23 \, \mathrm{h} \ 56 \,\mathrm{min})$ of a sidereal day, and $\phi$ a constant phase offset. Note that the function $\Delta \Phi_{\mathrm{corr}}^{\mathrm{(j)}}(t)$ is defined for each subrun separately whereas the sine function describes the possible Lorentz violating phase shift for all subruns. 


\section{Results and discussion}
The best combined fit of all 7 subruns with the parameter given by equation \refeq{aba:f=fi+sin+cos} results in a Lorentz violating phase shift of $\Phi_{\mathrm{LV}}=$(2.25$\,\pm\,$2.29)\,mrad (95\,$\%$ C.L.) corresponding to $\delta\nu=\Phi_{\mathrm{LV}}\cdot\Omega_{\mathrm{E}}/(2\pi)=(26.1\,\pm\,26.6) \, \mathrm{nHz}$. The correlated and the uncorrelated 1$\sigma$-errors of $\Phi_{\mathrm{LV}}$ are
\begin{eqnarray}
\Delta\Phi_{\mathrm {LV},\mathrm {correlated}}  &=& 1.01 \cdot 10^{-3} \, \mathrm{rad}, \notag \\
\Delta\Phi_{\mathrm {LV},\mathrm {uncorrelated}}&=& 1.84 \cdot 10^{-5} \, \mathrm{rad} ~.
\label{aba:errors}
\end{eqnarray}
The uncorrelated error is smaller than the correlated one by a factor of about 50. 
The phase amplitude $\Phi_{\mathrm{LV}}$ is mainly correlated with the fit parameters $E_{\mathrm{He(Xe)}}^{\mathrm{(j)}}$ from the exponential terms in Eq.\ \refeq{aba:modellorentzfit}, since the latter ones show a similar time structure as the sidereal modulation in the inital time intervals $\Delta T \approx 4$--$5 \,\mathrm{\ h}$ of each subrun where the SNR is still large.

The Lorentz violating SME coefficient $\tilde{b}_{\bot}^{\mathrm n}$ of the bound neutron\cite{kostelecky} can be calculated by
\begin{equation}
 \tilde{b}_{\bot}^{n}=\frac{\delta \nu \cdot h}{2 \cdot \sin\chi}\left(\frac{\gamma_{\mathrm{He}}}{\gamma_{\mathrm{Xe}}}-1\right)^{-1}\leq 3.72 \times 10^{-32} \, \mathrm{GeV} \ (95\,\% \ \mathrm{C.L.}),
\label{eq:b}
\end{equation}
where \textit{h} is the Planck constant, $\chi$ the angle between the Earth's rotation axis and the quantization axis of the spin ($\chi=57 ^\circ$), and $\gamma_{\mathrm{He}} / \gamma_{\mathrm{Xe}}\approx 2.75$.
 
Up to now, our experiments determine only an upper limit for the Lorentz violating SME coefficient, but we still have room for improvements. The increase of the $T_{2}^{*}$ time, and thus longer observation time \textit{T} reaching at least 24\,h for each subrun, would greatly reduce the correlated error. At present, the relatively short wall relaxation time of xenon ($8 \, \mathrm{h} < T_{1,\mathrm{wall}}^{\mathrm{Xe}} < 16\, \mathrm{h}$) limits the total observation time \textit{T} in our experiment. Therefore efforts to increase $T_{1,\mathrm{wall}}^{\mathrm{Xe}}$ considerably are essential. For helium $T_{1,\mathrm{wall}}^{\mathrm{He}} > 100 \ \mathrm{h}$ seems possible and transverse spin relaxation times $T_{1}$ of up to 60\,h have already been observed\cite{ourPaper}. Longer observation times $T$ also cause a higher sensitivity in frequency measurement according to the Cramer-Rao lower bound (Eq.\ \refeq{aba:crlb}). The exponential terms which are strongly correlated to the Lorentz violating amplitude $\Phi_{\mathrm{LV}}$ can be reduced considerably by rotating the cell to an appropriate position with respect to the magnetic guiding field. This way, one can minimize demagnetization induced effects\cite{ourPaper}.


\section*{Acknowledgments}

This work was supported by the Deutsche Forschungsgemeinschaft (DFG) under contract number BA 3605/1-1.

\end{document}